\documentclass[
aps, prd, reprint, notitlepage, a4paper,
floats, floatfix,
amsmath, amssymb, amsfonts, eqsecnum,
superscriptaddress,
showpacs, showkeys,
nofootinbib,
longbibliography,
]{revtex4-2}

\usepackage[left=20mm,right=20mm,top=25mm,bottom=25mm]{geometry}
\usepackage{graphicx} 
\usepackage[dvipsnames]{xcolor}
\usepackage{xspace} 
\usepackage{bm} 
\usepackage{amsmath,amsfonts,amssymb,amsthm}
\usepackage[utf8]{inputenc} 
\usepackage[normalem]{ulem}
\usepackage{overpic}
\usepackage{tikz}

\xdefinecolor{customcolor}{rgb}{0.0, 0.53, 0.74}
\usepackage[
	bookmarksnumbered, bookmarksopen, bookmarksopenlevel=2,
	breaklinks=true,
	colorlinks=true, filecolor=customcolor, citecolor=customcolor,
	linkcolor=customcolor, urlcolor=customcolor, menucolor=customcolor,
]{hyperref}

\newcommand{\icasu}{\affiliation{The Grainger College of Engineering, Illinois Center for Advanced Studies of the Universe \& \\ Department of Physics, University of Illinois Urbana-Champaign, Urbana, IL 61801, USA}}
\newcommand{\UNLVphysics}{\affiliation{Department of Physics and Astronomy, University of Nevada, Las Vegas, 4505 South Maryland Parkway, Las Vegas, NV 89154, USA}}
\newcommand{\NCfA}{\affiliation{Nevada Center for Astrophysics, University of Nevada, Las Vegas, NV 89154, USA}}
\newcommand{\milan}{\affiliation{Dipartimento di Fisica ``G. Occhialini'', 
Universit\`a degli Studi di Milano-Bicocca, Piazza della Scienza 3, 20126 Milano, Italy}}
\newcommand{\infn}{\affiliation{INFN, Sezione di Milano-Bicocca, 
Piazza della Scienza 3, 20126 Milano, Italy}}
\newcommand{\msu}{\affiliation{eXtreme Gravity Institute, Department of Physics,
Montana State University, Bozeman, MT 59717, USA}}

\begin{document}

\title{Uncertainty-aware waveform modeling for high signal-to-noise ratio gravitational-wave inference}

\hypersetup{pdftitle={Uncertainty-aware waveform modeling for high signal-to-noise ratio gravitational-wave inference}}

\author{Simone Mezzasoma}\email{simonem4@illinois.edu}\icasu
\author{Carl-Johan Haster}\email{carl.haster@unlv.edu}\UNLVphysics \NCfA
\author{Caroline B. Owen}\email{cbo4@illinois.edu} \milan\infn
\author{Neil J. Cornish}\email{ncornish@montana.edu}\msu
\author{Nicol\'as Yunes}\email{nyunes@illinois.edu}\icasu

\date{\today}

\begin{abstract}
Semi-analytical waveform models for black hole binaries require calibration against numerical relativity waveforms to accurately represent the late inspiral and merger, where analytical approximations fail. After the fitting coefficients contained in the model are optimized, they are typically held fixed when the model is used to infer astrophysical parameters from real gravitational-wave data. Though point estimates for the fitting parameters are adequate for most applications, they provide an incomplete description of the fit, as they do not account for either the quality of the fit or the intrinsic uncertainties in the numerical relativity data.
Using the IMRPhenomD~model, we illustrate how to propagate these uncertainties into the inference by sampling the fitting coefficients from a prior distribution and marginalizing over them. The prior distribution is constructed by ensuring that the model is compatible with a training set of numerical relativity surrogates, within a predefined mismatch threshold. This approach demonstrates a pathway to mitigate systematic bias in high signal-to-noise events, particularly when envisioned for the next generation of semi-analytical models.
\end{abstract}

\maketitle

\section{Introduction}
Fully analytical waveform models for gravitational waves (GWs) from binary black holes (BBHs) remain out of reach due to the complexity of solving the Einstein equations exactly. The most accurate waveforms we currently have to describe the late inspiral and merger of BBHs are obtained by evolving Einstein's equations numerically \cite{alcubierre, baumgarte2010numerical}. These numerical-relativity (NR) waveforms 
are computationally costly to produce \cite{Bruegmann:2006ulg, SpECwebsite,Loffler:2011ay}, and the limited NR data available renders them impractical for extensive data analysis. To address this limitation, faster semi-analytical waveform models have been constructed, primarily categorized into two families: phenomenological models \cite{Husa:2015iqa, Khan:2015jqa, Pratten:2020fqn, Pratten:2020ceb} and effective-one-body (EOB) models \cite{Bohe:2016gbl, Cotesta:2018fcv, Cotesta:2020qhw, Pompili:2023tna,Nagar:2018zoe,Nagar:2023zxh,Nagar:2024dzj,Nagar:2024oyk,Albanesi:2025txj}.
Additional effort to improve NR efficiency has led to NR surrogates \cite{Blackman:2015pia,Blackman:2017pcm,Varma:2018mmi}, which interpolate between known NR waveforms to generate high-quality waveforms in regions without NR simulations. However, NR surrogates are limited by the few NR GW cycles available, as generating the inspiral phase with NR methods is prohibitively expensive. Unless hybridized to post-Newtonian (PN) inspirals, NR surrogates are less suitable for parameter estimation of signals where the inspiral contributes significantly to the signal-to-noise ratio (SNR), as with low-mass BBHs and binary neutron stars.

We here focus on phenomenological models because they are particularly efficient for data analysis by computing closed-form expressions for GW polarizations directly in the frequency domain. The model amplitude and phase smoothly transition from the low-frequency inspiral regime through the merger to the high-frequency ringdown. Each segment uses a different analytical expression, tailored to the theoretical approximation that most reliably describes the dynamics in each regime. Despite recent advances in PN theory (see, e.g., \cite{Marchand:2016vox, Marchand:2017pir, Blanchet:2023sbv}), analytical methods alone are insufficient in the region leading up to (and including) the merger, requiring the use of higher-order corrections containing coefficients fit to NR data. The values of these coefficients are then traditionally held fixed during subsequent GW data inference with the model. 

Using point estimates for the fitting coefficients in phenomenological waveform models is sufficient if 
\begin{itemize}
    \item[(1)] the model accurately represents the NR data, i.e., the model is able to replicate the NR data throughout the parameter space with sufficient accuracy, and 
    \item[(2)] the biases in the inferred astrophysical parameters produced by the uncertainties in NR waveforms are undetectable. 
\end{itemize}
To address cases where condition (1) is not satisfied, model complexity can be increased. This includes strategies like adding more fitting coefficients (e.g. in refining \texttt{IMRPhenomD} into \texttt{IMRPhenomXAS} \cite{Pratten:2020fqn}), using a mixture model to capture any residual unfitted data, or selecting the optimal parametrization through transdimensional sampling \cite{Cornish:2015pda, Ellis:2016mtg, Littenberg:2020bxy, Gupta:2023jrn} (e.g. via Reversible Jump Markov Chain Monte Carlo~\cite{green}).  However, for GW detector data with sufficiently high SNRs, we become sensitive to the resolution of the NR waveforms \cite{Purrer:2019jcp}, and assumption (2) no longer holds.
At such SNRs, a model that can accurately reproduce NR waveforms should, in principle, be able to account for the inherent uncertainty in the NR waveforms as well. In this paper, we seek to propagate this uncertainty into the waveform, by allowing the model phenomenological fitting coefficients to vary within a prior volume, where the model faithfully reproduces NR data. We illustrate this approach using the phase of \texttt{IMRPhenomD} during inspiral, though this method can be applied to more recent models and can be generalized to also include merger and ringdown.

Our strategy builds on a series of efforts aimed at improving waveform models to make them more reliable for high-SNR GW parameter estimation. Recent studies have focused on quantifying waveform inaccuracies and incorporating them into GW inference through marginalization, similar to what is done in detector calibration \cite{Doctor:2017csx,Jan:2020bdz,Read:2023hkv,Bachhar:2024olc, Kumar:2025nojournal}. Notably, 4-PN-order truncation errors have been shown to introduce biases in inferred parameters at the sensitivities of current detector networks but can be marginalized away at the cost of increased statistical uncertainty \cite{Owen:2023mid}.

Probabilistic waveform models have also emerged as a means to incorporate uncertainty directly into waveform generation. These models employ Gaussian process regression (GPR) \cite{Moore:2014pda, Gair:2015nga, Moore:2015sza} to capture uncertainties in NR training waveforms, dynamically adjusting model precision based on the density and resolution of the training data \cite{Williams:2019vub}. While completing this project, we became aware of independent studies that employed GPR to represent NR uncertainty in both phenomenological \cite{Khan:2024whs} and EOB \cite{Bachhar:2024olc} waveform flavors. Additionally, EOB waveform uncertainty has been modeled with a mixture density network and marginalized over during inference \cite{Pompili:2024nojournal}.

Though similar in scope to probabilistic phenomenological models, our approach is more simple and direct, sampling the fitting coefficients from a probability distribution. We explicitly parametrize uncertainty at the level of the phase, following the approach of \cite{Read:2023hkv, Owen:2023mid}, by introducing frequency-dependent corrections. We can thus incrementally activate individual coefficients and regulate the transition from uninformative priors (i.e. likelihood-dominated constraint on the coefficients, as done in the training) to informative priors (i.e. prior-dominated constraint on the coefficients, as done in the GW inference). The training also reveals the underlying correlations among the coefficients, thus providing insight into the effectiveness of the parametrization: the most degenerate coefficients do not contribute to the waveform structure and could be reduced with a change of variable for a more efficient representation. 
The computational cost of sampling and marginalizing over the fitting coefficients remains comparable to that of the underlying waveform model, in this case, \texttt{IMRPhenomD}.

We organize the paper as follows. Section~\ref{sec:phenom-calibration} reviews the \texttt{IMRPhenomD} inspiral-phase calibration and the uncertainties from fitting to NR waveforms. Section~\ref{sec:training} details the construction of a probability distribution for fitting coefficients, ensuring consistency with NR surrogates. Section~\ref{sec:inference} applies this uncertainty-aware model in Bayesian inference, demonstrating improved parameter recovery over the standard \texttt{IMRPhenomD}. Section~\ref{sec:conclusions} summarizes findings and explores extensions, including dimensionality reduction for more efficient inference.

Throughout, we adopt geometric units, setting $G = c = 1$, so that all quantities can be expressed in seconds. For clarity, black hole masses are expressed in solar masses, with the unit conversion $1 \, M_\odot = 4.926 \times 10^{-6}\,\mathrm{s}$. To distinguish between time-domain and frequency-domain representations of the waveform model, we denote the time-domain waveform by \( h \) and its Fourier transform by \( \tilde{h} \).

\section{Phenom Model Calibration}
\label{sec:phenom-calibration}
In this section, we review the modeling of the inspiral phase in \texttt{IMRPhenomD} and examine the main sources of uncertainty associated with its calibration. We specifically focus on propagating uncertainties from the NR training data into the inference by introducing freedom in the higher-order phase corrections already built into the \texttt{IMRPhenomD} model.

\subsection{Inspiral phase}
\label{ssec:phase-inspiral}
\texttt{IMRPhenomD} \cite{Husa:2015iqa,Khan:2015jqa} is a phenomenological gravitational waveform model designed for quasi-circular BBH systems with spins aligned or anti-aligned along the orbital angular momentum. This model restricts attention to the dominant multipole in the positive frequency domain by specifying
\begin{equation}
    \tilde{h}_{+}(\boldsymbol{\theta}; f) = A(\boldsymbol{\theta}; f) e^{-i\phi(\boldsymbol{\theta};f)},
    \label{eq:hplus-def}
\end{equation}
where the real-valued amplitude \( A \) and phase \( \phi \) semi-analytically describe the (low-frequency) inspiral, intermediate, and (high-frequency) merger-ringdown regions. 
Here $\boldsymbol{\theta}$ denotes both intrinsic and extrinsic astrophysical parameters of the system. The intrinsic parameters include the component masses $m_{1,2}$ ($m_1 > m_2$) and the respective dimensionless spins $\chi_{1,2}$ (defined in Eq. (1) of ~\cite{Khan:2015jqa}). The extrinsic parameters include the luminosity distance $D_L$, the inclination angle $\iota$ in the source frame, and the time and phase at coalescence $t_c$, and $\phi_c$.
The dominant-multipole cross-polarization is uniquely determined by Eq.~\eqref{eq:hplus-def}, since
\begin{equation} \tilde{h}_{\times}(\boldsymbol{\theta}; f) = -2 i \frac{c_\iota}{1 + c_\iota^2}  \tilde{h}_{+}(\boldsymbol{\theta}; f),
\end{equation}
with $c_\iota = \cos \iota $ the cosine of the inclination angle.

We limit our analysis to the inspiral only, which is sufficient to show how our calibration of the phenomenological coefficients can be incorporated in parameter estimation. 
Within the inspiral, the Fourier transform of the known PN series (for a comprehensive introduction, see e.g. \cite{Maggiore:2007ulw, poisson2014gravity}) under the stationary phase approximation yields \( A \) and \( \phi \) in the form of a perturbative expansion in powers of the orbital velocity,
\begin{equation}
    v = (\pi M f)^{1/3},
\end{equation} 
where $M = m_1 + m_2$ is the total mass of the system.

The inspiral phase of the \texttt{IMRPhenomD} model 
\begin{equation}
\phi(\boldsymbol{\theta}, \boldsymbol{\lambda}; f) = \phi_{\mathrm{F2}}(\boldsymbol{\theta}; f) + \frac{3}{128\eta}v^{-5} \sum_{i=8}^{11} \kappa_i \,v^i,
\label{eq:inspiral-phase}
\end{equation}
is a function of both the astrophysical parameters $\boldsymbol{\theta}$ and a set of coefficients $\boldsymbol{\lambda}$ that are to be fit to NR data. 
Here, $\eta = m_1 m_2/M^2$ is the symmetric mass ratio, and
\begin{equation*}
\phi_\mathrm{F2} (\boldsymbol{\theta}; f) = 2\pi f t_c - \phi_c 
+ \frac{3}{128\eta}v^{-5}\sum_{i=0}^7 \varphi_i \,v^i,
\end{equation*}
is the TaylorF2 \cite{Damour:2001tu,Damour:2002kr,Arun:2004hn,Buonanno:2009zt} approximant, covering up to 3.5PN order in the non-spinning and spin-orbit sector, and up to 2PN order in the quadratic-in-spin sector. The constants $\lbrace\varphi_i\rbrace$ are closed-form functions of the intrinsic parameters\footnote{These coefficients are independent of $v$, except for $\varphi_5$ and $\varphi_6$, which contain terms proportional to $\log v$.}, whose explicit expressions can be found in Appendix B of \cite{Khan:2015jqa}.

To capture a realistic inspiral, it is essential to enrich the known, slowly converging \cite{Damour:2000zb,Damour:2002kr} terms with phenomenological higher-order corrections\footnote{This remains necessary even when we update the phase by including more informative PN results, like the latest 4PN order non-spinning dynamics \cite{Damour:2014jta,Bernard:2015njp,Damour:2017ced,Marchand:2017pir,Jaranowski:2015lha} and 4.5PN order non-linear tail contributions \cite{Marchand:2016vox,Messina:2018ghh,Nagar:2018plt,Pratten:2020fqn}. The perturbative series is always bound to introduce a truncation error, which is currently dealt with by calibrating it away.}.
Conventionally, these corrections are cast as a pseudo-PN structure, represented by the second term in Eq.~\eqref{eq:inspiral-phase}, which extends the series to 5.5 PN order. Each coefficient ($i = 8,9,10,11$)
\begin{equation}
    \kappa_i = \frac{128}{\pi^{(i-5)/3}(i-5)}\sigma_{i-7},
\end{equation}
is mapped to the physical parameter space via ($j=1,2,3,4$)
\begin{align}
\sigma_j = &\; \lambda^j_{00} + \lambda^j_{10} \eta \nonumber \\
&+ (\chi_\mathrm{PN} - 1) \left( \lambda^j_{01} + \lambda^j_{11} \eta + \lambda^j_{21} \eta^2 \right) \nonumber \\
&+ (\chi_\mathrm{PN} - 1)^2 \left( \lambda^j_{02} + \lambda^j_{12} \eta + \lambda^j_{22} \eta^2 \right) \nonumber \\
&+ (\chi_\mathrm{PN} - 1)^3 \left( \lambda^j_{03} + \lambda^j_{13} \eta + \lambda^j_{23} \eta^2 \right),
\label{eq:lambda_def}
\end{align}
a second-order polynomial in the symmetric mass ratio $\eta$ and a third-order polynomial \cite{Marsat:2014xea,Bohe:2015ana} in the spin parameter
\begin{equation}
\chi_\mathrm{PN} = \frac{q\, \chi_1 +\chi_2}{1+q} - \frac{38\eta}{113}(\chi_1 + \chi_2)\,,
\label{eq:chi_pn_def}
\end{equation}
where we have introduced the mass ratio $q = m_1/m_2$ ($\geq1$ by convention).
The polynomial structure in Eq.~\eqref{eq:lambda_def} was chosen to resemble the known coefficients $\lbrace\varphi_i\rbrace$, and the order in $\eta$ was selected to represent the features of the NR data within the desired accuracy goal.
In this paper, we collectively denote the dimensionless fitting coefficients for the inspiral phase as
\begin{equation}
    \boldsymbol{\lambda} = \lbrace \lambda^{j}_{00}, \lambda^{j}_{10}, \lambda^{j}_{01},\ldots, \lambda^j_{23}\rbrace\,,
\end{equation}
which are to be determined by calibrating the model against NR-based waveforms.

The amplitude in Eq.~\eqref{eq:hplus-def}, with its leading PN order term given by  
\begin{equation}
A(\boldsymbol{\theta}; f) = \sqrt{\frac{5}{96 \pi^{4/3}}} \, (1 + c_\iota^2) \, \frac{\mathcal{M}^{5/6}}{D_L} \, f^{-7/6},
\label{eq:leading-amplitude}
\end{equation}
where the chirp mass is defined as \(\mathcal{M} = M \eta^{3/5}\), follows a similar PN expansion as the phase (see Eqs.~(29)-(30) in \cite{Khan:2015jqa}) and involves additional fitting parameters. However, we exclude the amplitude from the current recalibration as it lies beyond the scope of this analysis.

\subsection{Propagation of uncertainties}
\label{ssec:phenom-calibration}
When the \texttt{IMRPhenomD} model was first introduced, the fitting coefficients were calibrated using 19 NR-hybrid waveforms that spanned a range of mass ratios \(q \in [1,18]\) and spin parameters \(\chi_{1,2} \in [-0.95, 0.98]\). These NR-hybrid waveforms were generated by tapering \texttt{SEOBv2} \cite{Taracchini:2013rva} inspirals to NR mergers (obtained with \texttt{SpEC} \cite{Ossokine:2013zga,Hemberger:2012jz,Szilagyi:2009qz,Boyle:2007ft,Mroue:2013xna} and \texttt{BAM} \cite{Bruegmann:2006ulg,Husa:2007hp}). The calibration was carried out hierarchically, by first constraining the non-spinning subspace with a one-dimensional fit over $\eta$, and then extending the fit to the $\chi_\mathrm{PN} \neq 0$ subspace.

More recently, the model was recalibrated \cite{Lam:2023oga} directly against 11 purely-NR waveforms from the Simulating eXtreme Spacetimes (SXS) catalog \cite{Boyle:2019kee}, and by simultaneously optimizing all fitting coefficients using gradient descent. This recalibration improved the model's fidelity to NR waveforms by up to 45\% compared to the original, highlighting the effect that the selection of NR training data and the type of fitting method can have on the model accuracy.

In both calibrations, the fitting procedure generated a point estimate for the fitting coefficients, optimized to achieve the best match between the \texttt{IMRPhenomD} model and the NR-based waveforms. Traditionally, once these coefficients are determined, they are held fixed during gravitational wave data inference, with their effective prior represented as a Dirac delta function. This choice fails to account for the inherent uncertainties in the NR data as well as uncertainties arising from the choice of the functional form of the model fit itself.

When fitting the \texttt{IMRPhenomD} model to NR-based waveforms, three primary sources of uncertainty come into play:

\begin{enumerate}
    \item[(i)] Intrinsic numerical error in NR waveforms: Each NR waveform is subject to various inherent uncertainties, including those that arise from discretization (grid size) \cite{Campanelli:2005dd,Bruegmann:2006ulg,Husa:2007hp,Vincent:2019qpd}, error propagation from initial data (e.g., constraint violation \cite{baumgarte2010numerical}), and the finite radius where gravitational waves are extracted \cite{Boyle:2007ft,Scheel:2008rj,Pfeiffer:2007yz}.
    
    \item[(ii)] Selection bias from finite number of NR waveforms in the training set: The available NR dataset is inherently limited in size due to computational cost, and the waveforms need to be selected carefully to efficiently represent the parameter space.
    
    \item[(iii)] Limitations in the functional form of the model: The \texttt{IMRPhenomD} model uses a piecewise parametric form, particularly relying on the PN expansions for the inspiral phase. While physically motivated, this approximation may not fully capture the NR details, especially at higher frequencies and for high-spin systems, causing the model to fit some regions of parameter space well, while underperforming in others.
\end{enumerate}

Not all of these sources of uncertainty are created equal.
The uncertainty in (ii) is managed by selecting a \textit{minimally sufficient set} of NR-based waveforms, where adding additional waveforms does not significantly improve the fit (as done e.g. in Sec.~IX of \cite{Khan:2015jqa}). A denser validation set is then used to ensure that the model provides an adequate fit across the entire parameter space.
Uncertainties (i) and (iii), however, are not typically accounted for in the point estimates of the model fitting coefficients.

While uncertainties in (ii) and (iii) set the ultimate limit on how well any parametric model can fit NR data, their relative impact for a fixed amount of analytical information, depends on the modeling approach.
For phenomenological models for example, the mass-averaged mismatch of \texttt{IMRPhenomXAS} against SXS NR waveforms shows improvements of one to two orders of magnitude over \texttt{IMRPhenomD} (see e.g. Fig.~15 of \cite{Pratten:2020ceb}). This improvement is attributed not only to a larger NR training set (652 waveforms instead of 19, enabling extrapolation to mass ratios up to $q \sim 1000$), but also to a more flexible parameterization that captures the waveform dependence in the symmetric mass ratio and the individual spins, instead of limiting the NR calibration to an effective spin parameter.

Similarly, recent improvements in spin-aligned EOB waveform models highlight the interplay between the choice of functional form for NR-calibrated parameters and the choice of NR waveforms used to fit them. The proper NR completion of the non-circular dynamics in the final cycles of the BBH lowers the maximum mismatch of \texttt{TEOBResumS} \cite{Nagar:2018zoe} against SXS NR waveforms from $\sim 10^{-2}$ (\texttt{TEOBResumS-v4.1.0}) to $\sim 10^{-3}$ (\texttt{TEOBResumS-v4.3.2}) \cite{Nagar:2023zxh} for the $l=|m|=2$ mode. Moreover, the recent update of \texttt{TEOBResumS-Dal\'i} \cite{Nagar:2024dzj,Nagar:2024oyk} for non-circularized BBHs shows that a more effective Pad\'e resummation scheme alone\footnote{The choice of resummation scheme affected the EOB potentials and the waveform amplitude corrections.} can lower the median mismatch against NR from $\sim 10^{-3}$ to $\sim 3 \times 10^{-4}$. This suggests that the choice of parametrization can sometimes take precedence over the amount of analytical  content (e.g. number of PN terms in the inspiral) included in the waveform, when the goal is to optimize the mismatch against the available NR simulations.

The search for an optimal analytical description that best matches spin-aligned NR waveforms is not the focus of this paper, and thus, it is beyond its scope. In particular we do not directly address the bias induced by the functional form (iii), but instead we aim at incorporating the uncertainty from (i), intrinsic to NR-based waveforms, into the model.
By introducing sampling freedom in the inspiral fitting coefficients, we can make the model more ``aware'' of the training set used for calibration. 

We note that, while we only focus on BBH inspirals, a similar NR-based fitting and marginalization strategy has already been applied to the inference of neutron star properties from GWs. In particular, deviations from equation-of-state-independent relations have been incorporated through calibration-like parameters within inspiral~\cite{Chatziioannou:2018vzf} and postmerger~\cite{Breschi:2021xrx,Breschi:2022nojournal} signals. This procedure allows to take advantage of quasi-universal relations during inference without causing bias.

The following section will detail the construction of the prior distribution, \( p(\boldsymbol{\lambda}) \), for the late-inspiral fitting coefficients associated with $\sigma_2$, $\sigma_3$, and $\sigma_4$, corresponding to the 4.5, 5, and 5.5 PN orders, respectively.

\section{Training and Fitting for the Calibration Parameters of IMRPhenomD}
\label{sec:training}
In this section, we describe the training waveforms and the likelihood function used for Bayesian coefficient calibration. We then detail the tools used to sample the distribution of fitting coefficients, \(p(\boldsymbol{\lambda})\), and the methods to validate it.

\subsection{Training waveforms}
To demonstrate our method, we use surrogate waveforms based on NR as training data, instead of directly employing NR waveforms. We choose this approach mainly for convenience, with the future prospect of applying this method to actual NR data once the proof of concept is validated.

Specifically, we adopt the time-domain \texttt{NRHybSur3dq8} model~\cite{Varma:2018mmi}, which is tailored for BBHs with aligned spins. We further extract the dominant $(\ell, |m|)=(2,2)$ multipole only, in parallel with the construction of \texttt{IMRPhenomD}. To obtain the frequency-domain representation, we first zero-pad each time-domain waveform by 2 seconds after the ringdown and apply a Tukey window with a 0.4-second roll-off to smooth the edges. We then apply the Fast Fourier Transform (FFT) algorithm.

The \texttt{NRHybSur3dq8} model achieves accuracy comparable to that of the original NR data for mass ratio \(1 \leq q \leq 8\) and spin magnitudes \(\lvert \chi_{1,2} \rvert \leq 0.8\). The mismatch between \texttt{NRHybSur3dq8} and NR is at most \(3 \times 10^{-4}\) across the total mass range \(2.25 M_\odot \leq M \leq 300 M_\odot\). Recall that the match is independent of the spectral
noise density of the instrument if this is constant, but it does depend on the number of cycles  inside the sensitivity band, and thus, on the total mass of the binary. 

We use this mismatch value as an approximate measure of the intrinsic uncertainty in the surrogate waveforms.
In our simplified approach, the mismatch serves as a prescribed error budget, and we do not model the individual sources of uncertainty separately.
Specifically, we do not distinguish the different sources of uncertainty arising from the inspiral hybridization, from the limited number of NR training waveforms selected by the greedy algorithm during the surrogate construction, or from the uncertainties inherited from the NR waveforms themselves. A more detailed analysis of these uncertainties would be better carried out by the developers of the surrogate models, who can systematically vary the model construction parameters and assess their impact by generating multiple model realizations.

We fix the total mass at \(M = 10\,M_\odot\) to provide long-duration inspirals, valuable for calibration. The remaining intrinsic parameters of the binary are defined as
\begin{equation}
\boldsymbol{\xi} = \{q, \chi_1, \chi_2\},
\end{equation}
and the parameter space where the \texttt{NRHybSur3dq8} is valid is denoted by
\begin{equation}
    \Xi = \lbrace \boldsymbol{\xi} : 1 \leq q \leq 8, \, \lvert \chi_{1,2} \rvert \leq 0.8 \rbrace .
\end{equation}

\subsection{Likelihood}
The natural method to calibrate the fitting coefficients and obtain a probability distribution for them is Bayesian inference.
In this framework, a fitting procedure yields a probability distribution that encodes both the degree of uncertainty and the correlations among the inferred parameters. This, however, requires a choice of likelihood function, which serves as a generative model for the data \cite{Hogg:2010nojournal}. Here, we motivate the choice of likelihood and describe the simplifying assumptions we use to carry out the model training.

To effectively condense the information from the entire parameter space $\Xi$ into a simplified phenomenological waveform model, we select a discrete representative subset $\mathcal{S} \subset \Xi$, the construction of which will be detailed later. The calibration is carried out for each $\boldsymbol{\xi} \in \mathcal{S}$ and must balance both local and global properties of the fit: (i) locally, at the waveform level for a fixed $\boldsymbol{\xi}$, and (ii) globally, by integrating information from different $\boldsymbol{\xi}$ points spanning the parameter space.

The single-waveform fit is performed in the frequency domain, where we optimize the coefficients that enter the phase linearly in Eq.~\eqref{eq:inspiral-phase} to minimize the residual
\begin{equation}
   \tilde{r}(\boldsymbol{\xi}, \boldsymbol{\lambda}) =   \tilde{h}^{\mathrm{train}}(\boldsymbol{\xi}) - \tilde{h}(\boldsymbol{\xi}, \boldsymbol{\lambda}).
\end{equation}
Here $\tilde{h}^{\mathrm{train}}(\boldsymbol{\xi})$ is the plus polarization of the training waveform with fixed $D_L = 100$ Mpc, $\iota = 0$, $M = 10\, M_\odot$, and
\begin{equation}
    \tilde{h}(\boldsymbol{\xi}, \boldsymbol{\lambda}) = \tilde{h}_{+}(\boldsymbol{\theta}; f) \big|_{D_L, \iota, M} ,
\end{equation}
is the plus polarization of \texttt{IMRPhenomD} in Eqs.~\eqref{eq:hplus-def}, computed at the same values of $D_L$, $\iota$, and $M$ as the training waveform. We sample over the fit coefficients $\boldsymbol{\lambda}$ contained within Eq.~\eqref{eq:inspiral-phase}.
For each $\boldsymbol{\xi}$, we marginalize over the physically-irrelevant time and phase difference, $t_c$ and $\phi_c$. Up to a multiplicative constant, the likelihood is chosen to be
\begin{equation}
\mathcal{L}_{\boldsymbol{\xi}}(\boldsymbol{\lambda}) = \int_{-T/2}^{T/2} \frac{d t_c}{T} \int_0^{2\pi} \frac{d \phi_c}{2\pi}\, e^{  -\frac{1}{2} \|  r(\boldsymbol{\xi}, \boldsymbol{\lambda}) \|_{\boldsymbol{\xi}}^2 } ,
\label{eq:marginalized_likelihood}
\end{equation}
where  $\|  r(\boldsymbol{\xi}, \boldsymbol{\lambda}) \|_{\boldsymbol{\xi}}^2 = \left( r(\boldsymbol{\xi}, \boldsymbol{\lambda}) | r(\boldsymbol{\xi}, \boldsymbol{\lambda}) \right)_{\boldsymbol{\xi}} $, and the inner product  $(\cdot|\cdot)_{\boldsymbol{\xi}}$ between two waveforms $h_1$ and $h_2$ is defined as
\begin{equation}
    (h_1|h_2)_{\boldsymbol{\xi}} = 4\,\text{Re} \int \frac{\tilde{h}_1(f)\tilde{h}_2^*(f)}{\mathcal{E}(\boldsymbol{\xi} , f)}\,df.
    \label{eq:inner-product}
\end{equation}
The function $\mathcal{E}(\boldsymbol{\xi} , f)$ represents the uncertainty associated with the point $\boldsymbol{\xi}$, and weighs the residual differently in each frequency bin while also providing an overall scaling for the single-waveform likelihood.
If we knew the intrinsic uncertainty associated with the training waveform at the point $\boldsymbol{\xi}$ (i.e.~the waveform variation among multiple simulations of the same BBH) we would be able to translate this into a weight function $\mathcal{E}(\boldsymbol{\xi} , f)$ that captures the allowed residual variation. 
For example, an error budget expressed as a phase \textit{noise} could be mapped to a weight function \(\mathcal{E}(\boldsymbol{\xi}, f)\), analogous to how detector noise is characterized through its power spectral density.
In addition to the weight function \(\mathcal{E}(\boldsymbol{\xi}, f)\), one could further augment the model to include a Gaussian-in-magnitude spline \cite{Kumar:2025nwb} to represent a small residual phase noise budget that cannot be replicated by the current  fitting coefficients.

The provision of a comprehensive error budget by the surrogate model community (that relies on NR input) would be highly valuable, especially in the form of a quantified uncertainty in the evolution of the amplitude and phase. Gaining a clear understanding of the latter would allow for a more granular parametrization and would avoid exclusively relying on mismatch as a proxy for uncertainty.

For simplicity, we will assume this weighting to be uniform across frequencies, namely $\mathcal{E}(\boldsymbol{\xi} , f) = \mathcal{E}(\boldsymbol{\xi})$, and driven only by the uncertainty of the training data. These choices reduce $\mathcal{E}(\boldsymbol{\xi} , f)$ to an overall scaling, whose value can be estimated by the mismatch criteria described in Sec.~\ref{sec:bayes-fit}. 
The reason for using a Gaussian-distributed residual, as in Eq.~\eqref{eq:marginalized_likelihood}, is twofold. First, because we will train the model at an SNR of order $\mathcal{O}(10^2)$ or more, Eq.~\eqref{eq:marginalized_likelihood} approaches a Gaussian distribution in the parameters $\boldsymbol{\lambda}$. For a given variance  $\mathcal{E}$, a Gaussian likelihood represents a conservative choice because it minimizes the Fisher information. Second, using Eq.~\eqref{eq:marginalized_likelihood} we can use known results from gravitational wave inference (see appendix G of~\cite{Chatziioannou:2017tdw}) 
that link the variance of the posterior distribution to that of the waveform mismatch.

The global fit is performed through the global training likelihood
\begin{equation}
\mathcal{L}^{\mathrm{train}}(\boldsymbol{\lambda}) = \prod_{\boldsymbol{\xi} \in \mathcal{S}} \mathcal{L}_{\boldsymbol{\xi}}(\boldsymbol{\lambda}),
\label{eq:likelihood_train}
\end{equation}
defined by combining the contributions from the training waveforms in a chosen training set $\mathcal{S}$. In theory, one would construct the set  $\mathcal{S}$ hierarchically, as done in reduced order modeling through a greedy algorithm \cite{Smith:2016qas, Varma:2018mmi}. However, we simplify the template placement by selecting a uniform grid in the parameter space $\Xi$ and manually adding midpoints in higher-mismatch regions until the model saturates, i.e.~until adding new training waveforms no longer improves the global fit.

\subsection{Bayesian fit and validation}
\label{sec:bayes-fit}
This subsection details the sampling of the distribution, \( p(\boldsymbol{\lambda}) \), for the fitting coefficients associated with \(\sigma_{2}\), \(\sigma_{3}\), and \(\sigma_{4}\), corresponding to the 4.5, 5, and 5.5 PN orders respectively. The \(\sigma_{1}\) coefficient is kept as originally calibrated since most of the 4PN terms are currently known, and this term is of the same order as the \(2 \pi f t_c\) term (in the TaylorF2 phase), which we marginalize over. Each $\sigma_j$ is constructed from 11 coefficients $\lambda^{j}_{mn}$, as shown in Eq. \eqref{eq:lambda_def}. Therefore, we aim to sample 33 fitting coefficients, in total. Henceforth, we define 
\begin{equation}
\dim(\boldsymbol{\lambda}) = 33,    
\end{equation}
to represent the dimensionality of the coefficient vector $\boldsymbol{\lambda}$.

The distribution is obtained through Bayes' theorem~\cite{Gelman1995}
\begin{equation}
p(\boldsymbol{\lambda}) \propto \mathcal{L}^{\mathrm{train}}(\boldsymbol{\lambda}) \, \pi(\boldsymbol{\lambda}),
\label{eq:posterior_lambda_bayes}
\end{equation}
after supplying the likelihood in Eq.~\eqref{eq:likelihood_train}, and a prior $ \pi(\boldsymbol{\lambda})$. Given that we have no prior knowledge on the fitting coefficients other than their recently calibrated point-estimates, we assume the prior to be uniform and centered around the original calibration value. The width of the prior is chosen, so the posterior distribution remains well-supported within the bounds of the prior, rather than peaking at its edges.

To proceed, we consider the frequency range $[f_\mathrm{min}, f_\mathrm{max}]=[30, 366]$ Hz. The lower bound is chosen to limit the waveform length, so a sufficient number of training waveforms can be loaded into the available memory resources. When the total mass is $M = 10 M_\odot$, $ M f_\mathrm{min}= 0.0015$, which is already below the limit of validity $M f_\mathrm{min} = 0.0035$ for the original \texttt{IMRPhenomD} model. Furthermore, the beyond-4-PN-order terms we aim to constrain are not expected to have a significant impact at low frequencies, so lowering the frequency bound further is unnecessary.
The upper bound, defined by 
$M f_\mathrm{max} = 0.018$, corresponds to the transition from the inspiral to the 
intermediate phase in the \texttt{IMRPhenomD} model. Because we assume a frequency-independent variance, $\mathcal{E}(\boldsymbol{\xi})$, the number of GW cycles effectively used in the training is solely determined by the frequency range $[f_\mathrm{min}, f_\mathrm{max}]$.

The variability in the fitting coefficients is constrained by the requirement that the trained model remains \textit{indistinguishable} (see e.g. \cite{Toubiana:2024nojournal} and references therein) from the training data within a training SNR. Specifically, we require that by sampling $\boldsymbol{\lambda}$ from $p(\boldsymbol{\lambda})$, the mismatch between each training waveform and the model
\begin{equation}
\text{MM}_{\boldsymbol{\xi}}(\boldsymbol{\lambda}) = 1 - \max_{t_c, \phi_c}  \frac{(h^{\mathrm{train}}(\boldsymbol{\xi}) | h(\boldsymbol{\xi}, \boldsymbol{\lambda}))_{\boldsymbol{\xi}}}{\|  h^{\mathrm{train}}(\boldsymbol{\xi}) \|_{\boldsymbol{\xi}}  \| h(\boldsymbol{\xi}, \boldsymbol{\lambda}) \|_{\boldsymbol{\xi}}}  ,
\label{eq:mismatch-def}
\end{equation}
yields a distribution \(p(\text{MM}_{\boldsymbol{\xi}}(\boldsymbol{\lambda}))\) with a predefined variance \(\sigma_{\text{MM}}^2\).
The value set for $\sigma_{\text{MM}}$ should be no smaller than the intrinsic mismatch in each training waveform (computed, for example, by comparing the training waveform at its highest and second-highest numerical resolution). For simplicity, we assume this intrinsic mismatch to be the same for each point $\boldsymbol{\xi}$. Hence, we set $\sigma_{\text{MM}} = 10^{-4}$ to mimic the accuracy of the \texttt{NRHybSur3dq8} model against NR.

Selecting a value for $\sigma_{\text{MM}}$ fixes the total SNR of the training waveforms to \cite{Chatziioannou:2017tdw}
\begin{equation}
\rho^\mathrm{train}_\mathrm{tot} = \sqrt{\frac{\mathrm{dim(\boldsymbol{\lambda})-1}}{\sqrt{2} \, \sigma_{\text{MM}}}} .
\end{equation}
For $\sigma_{\text{MM}} = 10^{-4}$  and $\mathrm{dim(\boldsymbol{\lambda})}=33$, we obtain a training SNR of $\rho^\mathrm{train}_\mathrm{tot} = 475.7$, 
which we equally partition among the training set, so each point in parameter space carries an equal weight in the fit. Namely, we scale the value of $\mathcal{E}(\boldsymbol{\xi})$ in each inner product in Eq.~\eqref{eq:inner-product} so that
\begin{equation}
     \|  h^{\mathrm{train}}(\boldsymbol{\xi}) \|_{\boldsymbol{\xi}} = \frac{\rho^\mathrm{train}_\mathrm{tot}}{\sqrt{N}} \qquad \forall \boldsymbol{\xi} \in \mathcal{S},
\end{equation}
where $N$ is the number of waveforms in the set \(\mathcal{S}\).

To construct \(\mathcal{S}\), we begin by generating a uniform grid of $4\times 4\times 4$ points across the parameter space $\lbrace \boldsymbol{\xi} : 1 \leq q \leq 7.95, \, \lvert \chi_{1,2} \rvert \leq 0.79 \rbrace$. After running an initial training on these 64 points, we identify regions where the mismatch is suboptimal and add 30 additional points in those areas. For the selected mismatch threshold \(\sigma_{\text{MM}} = 10^{-4}\), including further training points does not yield significant improvement so we finalize the training set at $N=94$ waveforms.

To perform the sampling of \(p(\boldsymbol{\lambda})\) in Eq.~\eqref{eq:posterior_lambda_bayes}, we use \texttt{flowMC} \cite{flowmc_paper}, a Python-based Markov chain Monte Carlo (MCMC)~\cite{Metropolis,Hastings:1970aa,Gelfand01061990,Robert2004,Gamerman2006} library that leverages a normalizing flow (NF) \cite{JMLR:v22:19-1028,Kobyzev2021} and uses automatic differentiation~\cite{Baydin:2015nojournal,margossian2018} through \texttt{jax}~\cite{jax2018github} for gradient computations.
\texttt{flowMC} proceeds in two phases: a \textit{learning phase}, where the NF adapts to approximate the target posterior distribution, and a \textit{production run}, where the trained NF is used to propose MCMC jumps.
The sampling is carried out with the Metropolis-adjusted Langevin algorithm~\cite{MALA_paper1996} on an NVIDIA A100 (80GB) GPU.
The \texttt{IMRPhenomD} waveform model is accessed through the \texttt{ripplegw} \cite{ripple_paper} library, a \texttt{jax}-compatible Python package that also benefits from auto differentiation.

As is clear from the parametrization in Eq.~\eqref{eq:inspiral-phase}, the fitting coefficients appear as a linear combination whose components are not orthogonal in either frequency space or among the training waveforms in the set $\mathcal{S}$. Consequently, the fitting coefficients that maximize the training likelihood are expected to be degenerate. This degeneracy, combined with the high dimensionality of the fit, makes it less efficient for the MCMC to produce statistically independent samples.
To aid convergence, we apply a linear transformation 
\begin{equation}
    \lambda'_i = \sum_{j = 1}^{\dim(\boldsymbol{\lambda})} P_{ij} \lambda_j ,
    \label{eq:forward-transformation}
\end{equation}
to weaken these correlations, and sample in the transformed parameter space, $\boldsymbol{\lambda}'$.
We define $P_{ij}$ as 
\begin{equation}
    P_{ij} = [\boldsymbol{e}_1| \boldsymbol{e}_2| \ldots| \boldsymbol{e}_{\dim(\boldsymbol{\lambda})}]_{ij},
\end{equation}
where $\lbrace\boldsymbol{e}_i\rbrace_i$ are the unit-eigenvectors of the observed Fisher information matrix
\begin{equation}
F_{ij} (\boldsymbol{\lambda}) = - \frac{\partial^2}{\partial \lambda_i \partial \lambda_j} \mathcal{L}^{\mathrm{train}} (\boldsymbol{\lambda}),
\end{equation}
evaluated at the sampler starting point, $\boldsymbol{\lambda}_\mathrm{start}$, which we initially take to be the original \texttt{IMRPhenomD} calibration. 

\begin{figure}
    \centering
    \includegraphics[width=\columnwidth]{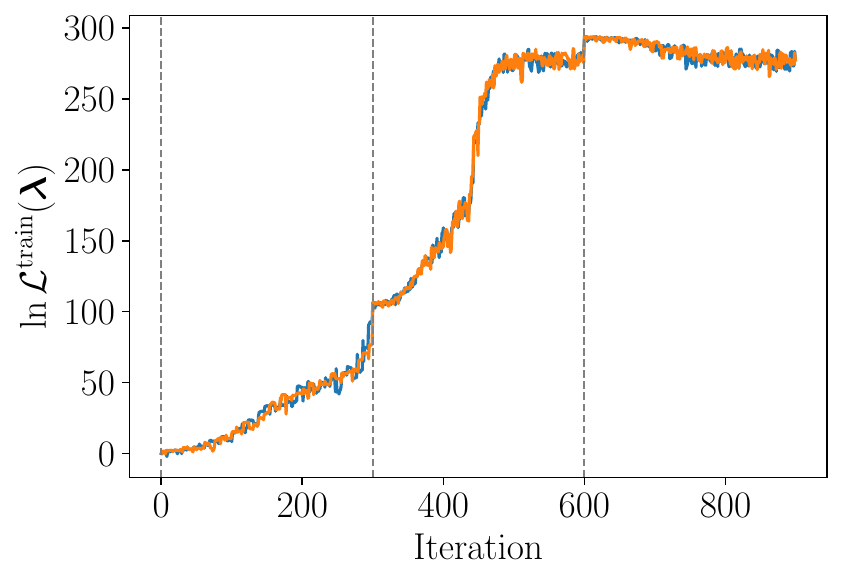}
    \caption{Log-likelihood values along two thinned chains during the learning phase of \texttt{flowMC}. The vertical dashed lines denote the injection point and the following two checkpoints. The saturation of the likelihood values in the last segment of the learning phase, combined with other convergence diagnostics~\cite{Roy2020} (Gelman-Rubin statistic~\cite{Gelman:1992zz,Brooks} below $1.05$ and good chain mixing) provide strong evidence of global convergence.}
    \label{fig:loglikelihood_along_train_chain}
\end{figure}

Figure~\ref{fig:loglikelihood_along_train_chain} shows the values of the likelihood in Eq.~\eqref{eq:likelihood_train} along two parallel chains as they find and populate the maximum during the learning phase of \texttt{flowMC}.
The search is segmented into checkpoints to avoid saturating the GPU memory, while focusing on the  latest explored region of the parameter space. At each checkpoint, $\boldsymbol{\lambda}_\mathrm{start}$ is updated to the mean of the most recent 100 samples, and the linear transformation in Eq.~\eqref{eq:forward-transformation} is recomputed.
The likelihood spikes at each reinjection because the mean of the last MCMC samples gives a better estimate of the maximum likelihood point compared to the single highest-likelihood value encountered.
With each reevaluation, the transformation becomes increasingly more effective at decorrelating the parameters, as the chains converge toward the high-probability region.
After the learning phase, the NF becomes stationary and is used to propose samples during the production run. 
We recover the original parameters post-sampling 
\begin{equation}
    \lambda_i = \sum_{j = 1}^{\dim(\boldsymbol{\lambda})} P_{ji} \lambda_j',
\end{equation}
by applying the transpose\footnote{This makes the inversion numerically stable, even when $P_{ij}$ is ill-conditioned.} of the most recently injected $P_{ij}$, whose orthogonality follows from $F_{ij}$ being real and symmetric.

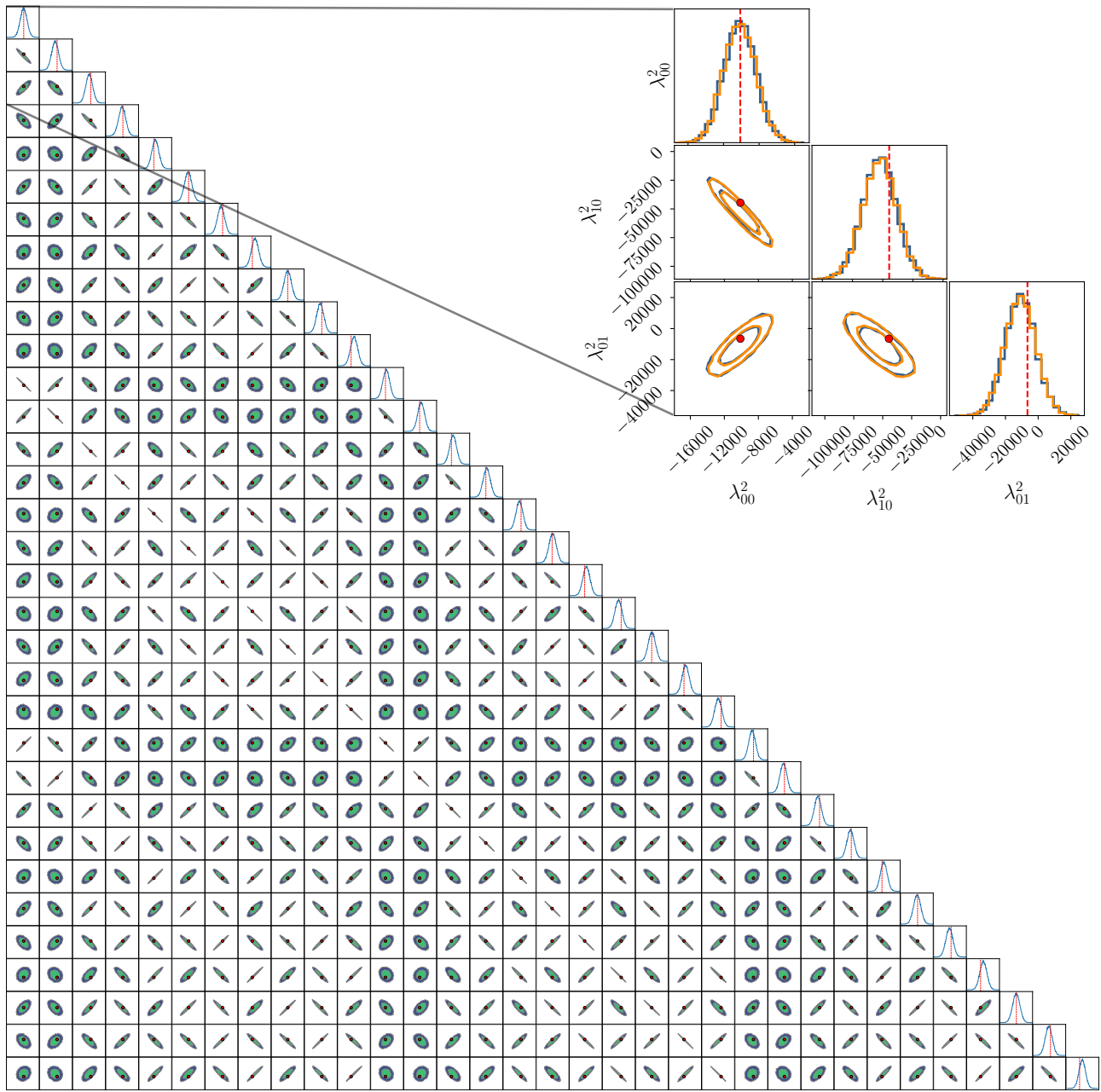
\begin{figure*}
    \centering
    \begin{overpic}[width=\textwidth]{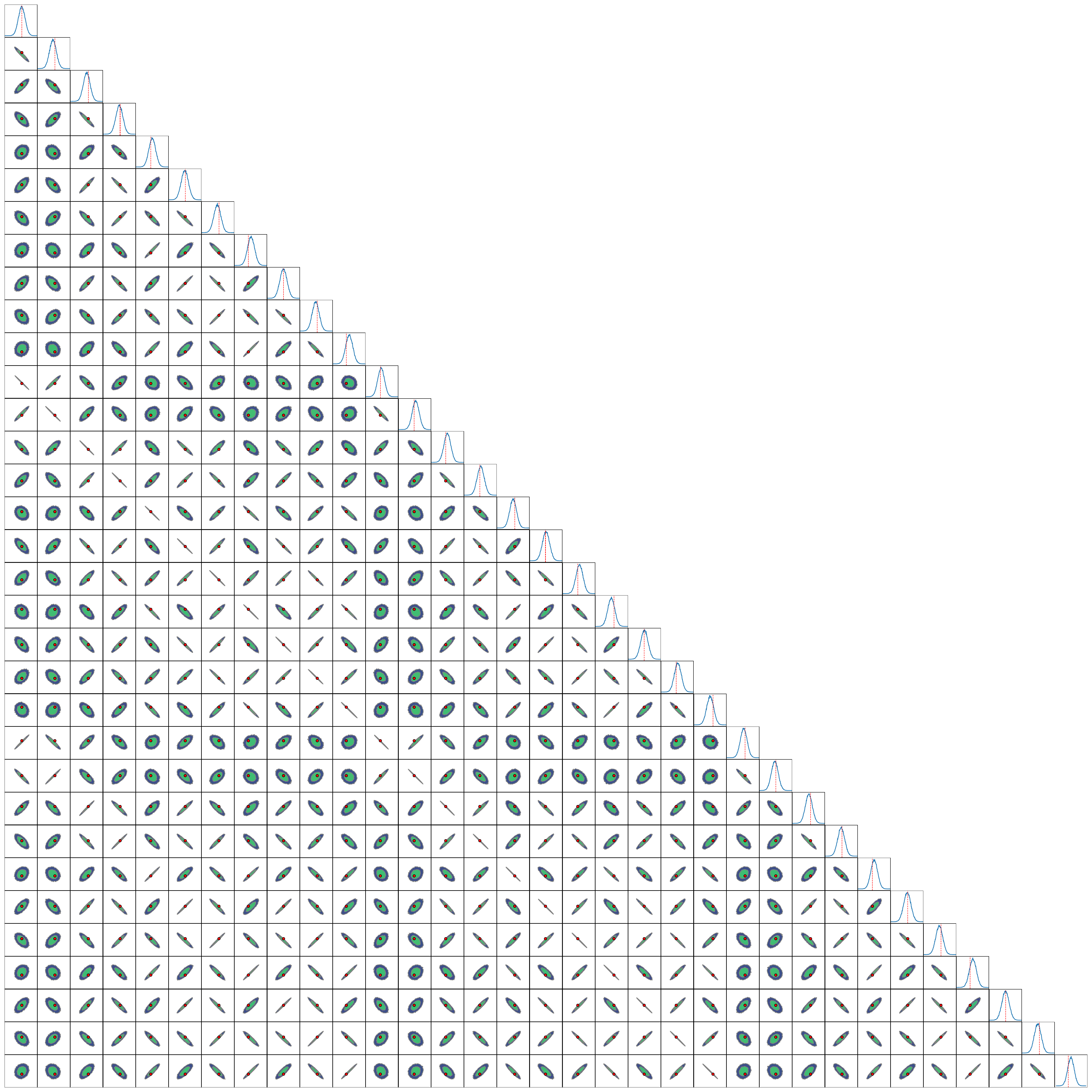}
        \put(51,52.7){
            \includegraphics[width=\columnwidth]{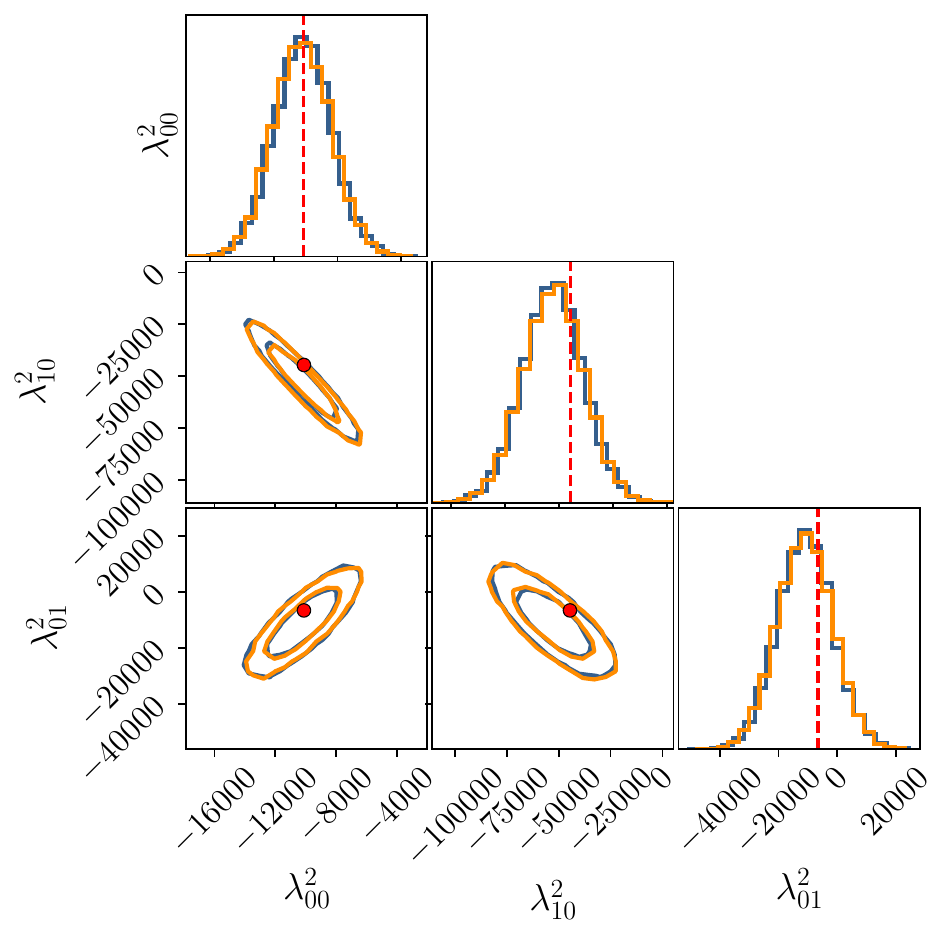}
        }
        \begin{tikzpicture}[overlay, remember picture]
            \draw[line width=1pt, opacity=0.5] (0.08, 15.4) -- (10.36, 10.57);
            \draw[line width=1pt, opacity=0.5] (0.08, 16.92) -- (10.36, 16.88);
        \end{tikzpicture}
        
    \end{overpic}
    \caption{Corner plot showing two-dimensional projections of the probability distribution $p(\boldsymbol{\lambda})$ for the 33 fitting coefficients $\boldsymbol{\lambda} = \{\lambda_{00}^2, \lambda_{10}^2, \lambda_{01}^2, \dots, \lambda_{23}^2, \lambda_{00}^3, \dots, \lambda_{23}^4\}$. The contours enclose the 60\% and 90\% highest density regions (HDRs). The red dots and lines mark the original \texttt{IMRPhenomD} values for the fitting coefficients, as implemented in \texttt{ripplegw}. Observe that the original calibration values are supported by the probability distribution \( p(\boldsymbol{\lambda}) \), as they fall within the 90\% HDR. The major degeneracies arise between coefficients carrying the same mass and spin dependence but entering at different PN orders, e.g. between $\lambda_{00}^2, \lambda_{00}^3$, and $\lambda_{00}^4$.
    The inset in the top right displays the first three dimensions of the probability distribution \( p(\boldsymbol{\lambda}) \) (blue), overlaid with a multivariate Gaussian distribution (orange) constructed using the mean and covariance of \( p(\boldsymbol{\lambda}) \). Observe that \( p(\boldsymbol{\lambda}) \) is well represented by a multivariate Gaussian.
}
    \label{fig:flowmc_full_corner}
\end{figure*}

The samples resulting from the production run are shown in the large corner plot of Fig.~\ref{fig:flowmc_full_corner}. The two-dimensional marginal distributions reveal strong correlations among the coefficients that multiply the same mass and spin combinations, but that enter at different PN orders. This suggests that the effects of the individual PN phase corrections cannot be resolved by the training data, at least when considering inspiral-only waveforms. This is in line with previous studies of astrophysical parameter estimation with \texttt{IMRPhenomD}, which showed that there can be bias-inducing ``cross-talk'' between phase terms that are 0.5PN order apart~\cite{Owen:2023mid}.

Furthermore, a visual inspection of Fig.~\ref{fig:flowmc_full_corner} shows that the samples follow a multivariate Gaussian distribution. To quantify the agreement with a multivariate Gaussian, we compare the one-dimensional marginal distributions of the samples to those of a Gaussian distribution constructed from the sample mean and covariance.
We do so by computing the Jensen–Shannon (JS) divergence \cite{MENENDEZ1997307} between these distributions and find values below $10^{-3}$ bits across all 33 dimensions\footnote{By definition, the JS divergence tends to zero when the two sets of samples are drawn from the same underlying distribution.}.

\begin{figure*}
    \centering
\includegraphics[width=\textwidth]{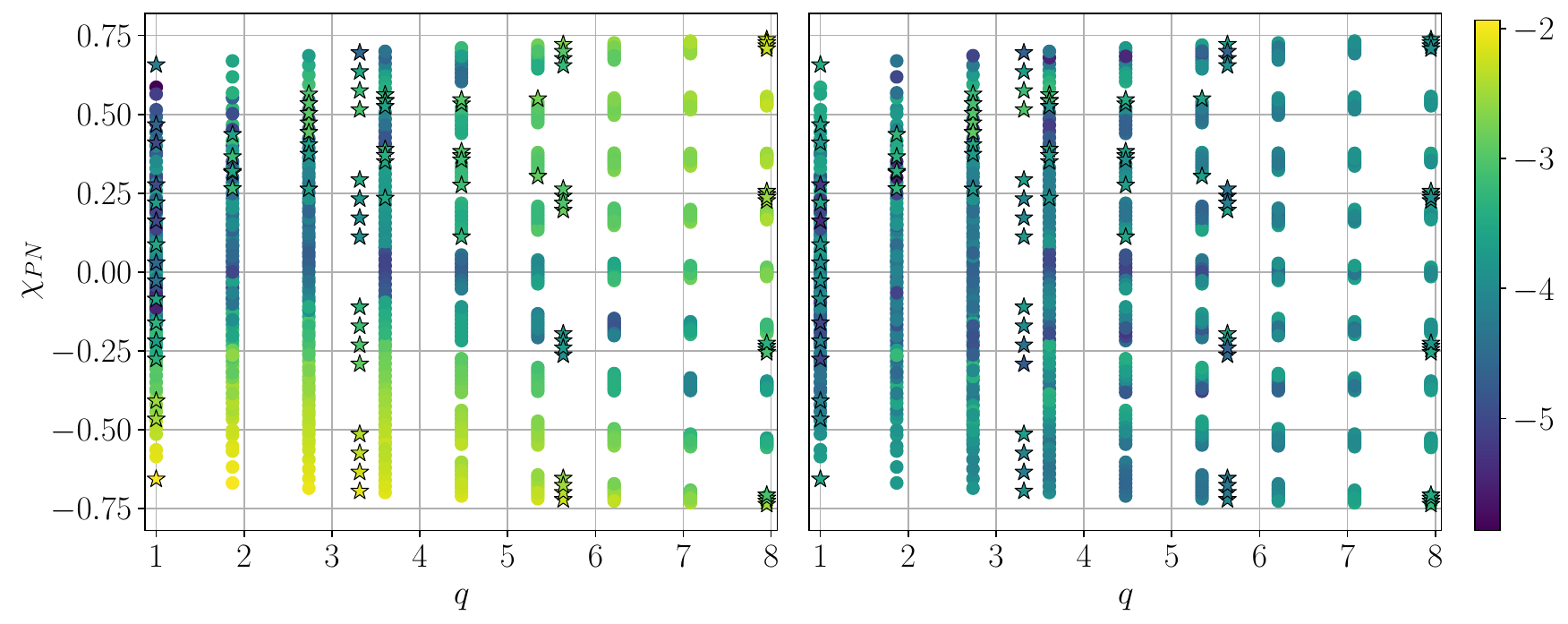}
    \caption{$\log_{10}$ mismatch of \texttt{IMRPhenomD} against the validation set of \texttt{NRHybSur3dq8} waveforms before (left) and after (right) the training. The post-training mismatches are computed at the mean of the distribution $p(\boldsymbol{\lambda})$. Observe that, before training, the mismatch spans three orders of magnitude, with values reaching $10^{-2}$ in regions of parameter space with small mass ratios ($q \lesssim 2$) and both spins negatively aligned ($\chi_\mathrm{PN} \lesssim -0.6$). After training, the mismatch improves consistently across the entire parameter space, with values uniformly reduced to  $\mathcal{O}(10^{-4})$. Star-shaped markers indicate the locations of the original 94-waveform training set, which is included in the validation set. The results confirm that the training successfully minimized the mismatch and that the 94-waveform set was sufficient for the model to generalize beyond the training points.
 }
 \label{fig:mismatch_validation}
\end{figure*}

To validate our training, we perform two checks. 
First, we verify that the mean of the distribution obtained from the training also minimizes the mismatch, as defined in Eq.~\eqref{eq:mismatch-def}, against each training waveform.
Figure~\ref{fig:mismatch_validation} compares the mismatch before (i.e., the default \texttt{IMRPhenomD}) and after the training, across the two-dimensional projection of $\Xi$ defined by the plane $(q, \chi_\mathrm{PN})$. Observe that, after training, the mismatch improves uniformly across the entire parameter space, a result of assigning equal weights to all training points in the fit. This indicates that (i) the likelihood in Eq.~\eqref{eq:likelihood_train} correctly finds the minimum mismatch, despite marginalizing over $t_c$ and $\phi_c$, rather than maximizing over them, and (ii) the 94-waveform training set is sufficiently dense for the model to interpolate regions of parameter space not included in the training.

\begin{figure}
    \centering
\includegraphics[width=\columnwidth]{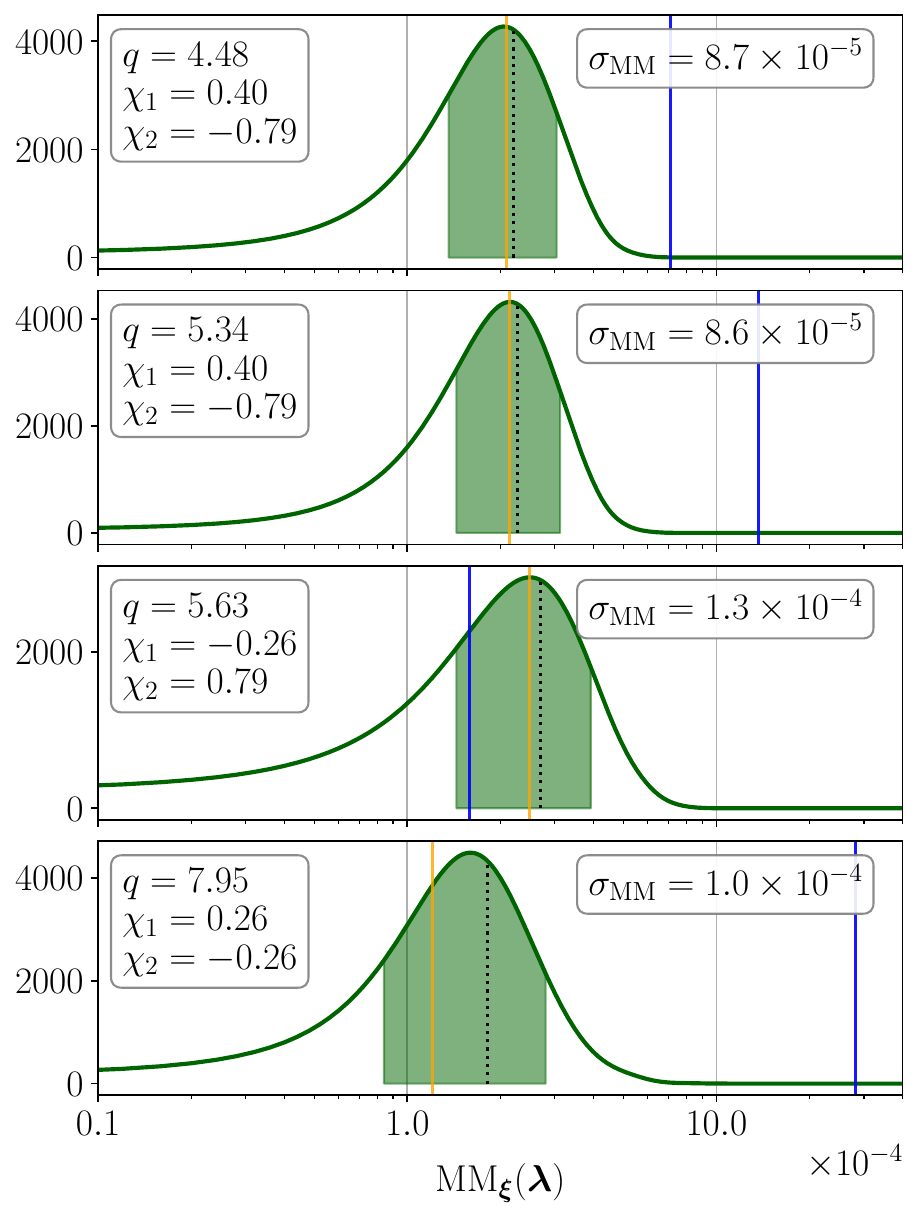}
    \caption{Mismatch distribution \( p(\mathrm{MM}_{\boldsymbol{\xi}}(\boldsymbol{\lambda})) \) with \(\boldsymbol{\lambda}\) sampled from \( p(\boldsymbol{\lambda})\). Four representative cases are shown, corresponding to  \(\boldsymbol{\xi} =  \lbrace 4.48, 0.40, -0.79 \rbrace\), \(\lbrace 5.34, 0.40, -0.79 \rbrace\), \(\lbrace 5.63, -0.26, 0.79 \rbrace\), \(\lbrace 7.95, 0.26, -0.26 \rbrace\), in order from top to bottom. The blue vertical lines denote the mismatch before the training. The yellow line indicates the mismatch at the mean of the distribution \( p(\boldsymbol{\lambda}) \), while the dotted black line marks the mismatch median. The green shaded region represents the interval corresponding to one standard deviation from the median. The standard deviation of each distribution successfully converged near its target value $\sigma_\mathrm{MM} = 10^{-4}$.}
    \label{fig:mismatch_distribution}
\end{figure}

As a second check, we confirm that the variance of the mismatch distribution  \(p(\text{MM}_{\boldsymbol{\xi}}(\boldsymbol{\lambda}))\) of the model against the training data converged to the target value $\sigma_\mathrm{MM}^2$. 
Figure~\ref{fig:mismatch_distribution} illustrates four (of the 94) mismatch distributions, highlighting how varying the fitting coefficients within \(p(\boldsymbol{\lambda})\) causes the model to differ from each training waveform, while maintaining a precision comparable to that of the training waveform itself (here quantified by the mismatch value $\sigma_{\text{MM}} = 10^{-4}$). In each panel of Fig.~\ref{fig:mismatch_distribution}, the blue vertical line indicates the mismatch before the training. In cases where the pre-training mismatch is already of order $10^{-4}$ or smaller, the post-training median mismatch can be slightly larger than the pre-training value, although it remains of the same order of magnitude. This results from our optimization of the global likelihood, where the accuracy on any individual waveform can be traded to improve the fit over the training set as a whole. Overall, nonetheless, observe that the mean mismatch after training can be more than one order of magnitude better than before training.

\section{GW Inference with uncertainty-aware IMRPhenomD}
\label{sec:inference}

With the validated distribution \( p(\boldsymbol{\lambda}) \) from the previous section, we can now incorporate the uncertainties of the fitting coefficients into parameter estimation of GW signals. 
More concretely, this section expands the parameter space we sample during the inference of GW data by relaxing the prior on the fitting coefficients from a Dirac delta function, centered at the best-fit value, to the multivariate Gaussian, $p_\mathrm{MVG}(\boldsymbol{\lambda})$, that best approximates $p(\boldsymbol{\lambda})$.
This section then present the results of applying this modified inference method to synthetic GW injections.

In our study we consider zero-noise \texttt{NRHybSur3dq8} injections, with strain data $D$ from the three-detector network consisting of the LIGO Hanford, LIGO Livingston \cite{LIGOScientific:2014pky}, and the Virgo \cite{VIRGO:2014yos} detectors. 
We assume that these detectors operate at design sensitivity and employ the O4 advanced LIGO \cite{LIGO-T2000012} and BNS-optimized advanced Virgo \cite{LIGO-P1200087_Virgo_PSD,KAGRA:2013rdx} (power spectral density) noise curves\footnote{The sensitivities can be found at \url{https://dcc.ligo.org/public/0165/T2000012/002/aligo_O4high.txt} and \url{https://dcc.ligo.org/public/0094/P1200087/042/fig1_adv_sensitivity.txt}}.

As in the training of Sec.~\ref{sec:training}, we evaluate our uncertainty-aware \texttt{IMRPhenomD} model through \texttt{ripplegw} to accommodate the simultaneous sampling of astrophysical parameters $\boldsymbol{\theta}$ and fitting coefficients $\boldsymbol{\lambda}$. 
The frequency range for the inference is set to $[f_\mathrm{min}, f_\mathrm{max}] = [20, 0.018/(M_{\mathrm{inj}}/\mathrm{s})]$ Hz, where $M_{\mathrm{inj}}$ represents the injected total mass. The upper bound is selected to crop the inspiral only, and thus, maintain consistency with the training. 

Parameter estimation is performed using the \texttt{bilby} \cite{bilby_paper} library and its parallelized version, \texttt{parallel-bilby} \cite{pbilby_paper}. We employ the \texttt{dynesty} \cite{2020MNRAS.493.3132S,sergey_koposov_2024_12537467} nested sampling algorithm within \texttt{bilby} for efficient exploration of the parameter space.
We marginalize over the luminosity distance ($D_L$) using a lookup table~\cite{Singer:2015ema,Thrane:2018qnx}. We also marginalize over the time of coalescence \(t_c\) using the FFT method~\cite{Farr:2014FFT}, and we marginalize over the phase of coalescence \(\phi_c\) analytically~\cite{Veitchetal:2013phase}.
Thus, we are left with the active sampling of the astrophysical parameters
\begin{equation}
    \boldsymbol{\theta} = \lbrace \mathcal{M}, q, \chi_1, \chi_2, \mathrm{dec}, \mathrm{ra}, \theta_{JN}, \psi\rbrace,
\end{equation}
where $\mathcal{M}$ is the chirp mass, \(\mathrm{dec}\) and \(\mathrm{ra}\) represent the declination and right ascension (sky localization), \(\theta_{JN}\) is the angle between the binary's total angular momentum and the line of sight, and \(\psi\) is the polarization angle (see Table E1 in \cite{Romero-Shaw:2020owr} for definitions).
The main output of the inference process is a posterior distribution $p(\boldsymbol{\theta}, \boldsymbol{\lambda} | D)$, from which we derive the probability density on astrophysical parameters alone by ``integrating out'' or marginalizing over the fitting coefficients
\begin{equation}
p(\boldsymbol{\theta}| D) = \int  p(\boldsymbol{\theta}, \boldsymbol{\lambda} | D) \,\mathrm{d}\boldsymbol{\lambda} .
\label{eq:marginalized_astro_distribution}
\end{equation}

\begin{figure*}
    \centering
    \includegraphics[width=0.9\textwidth]{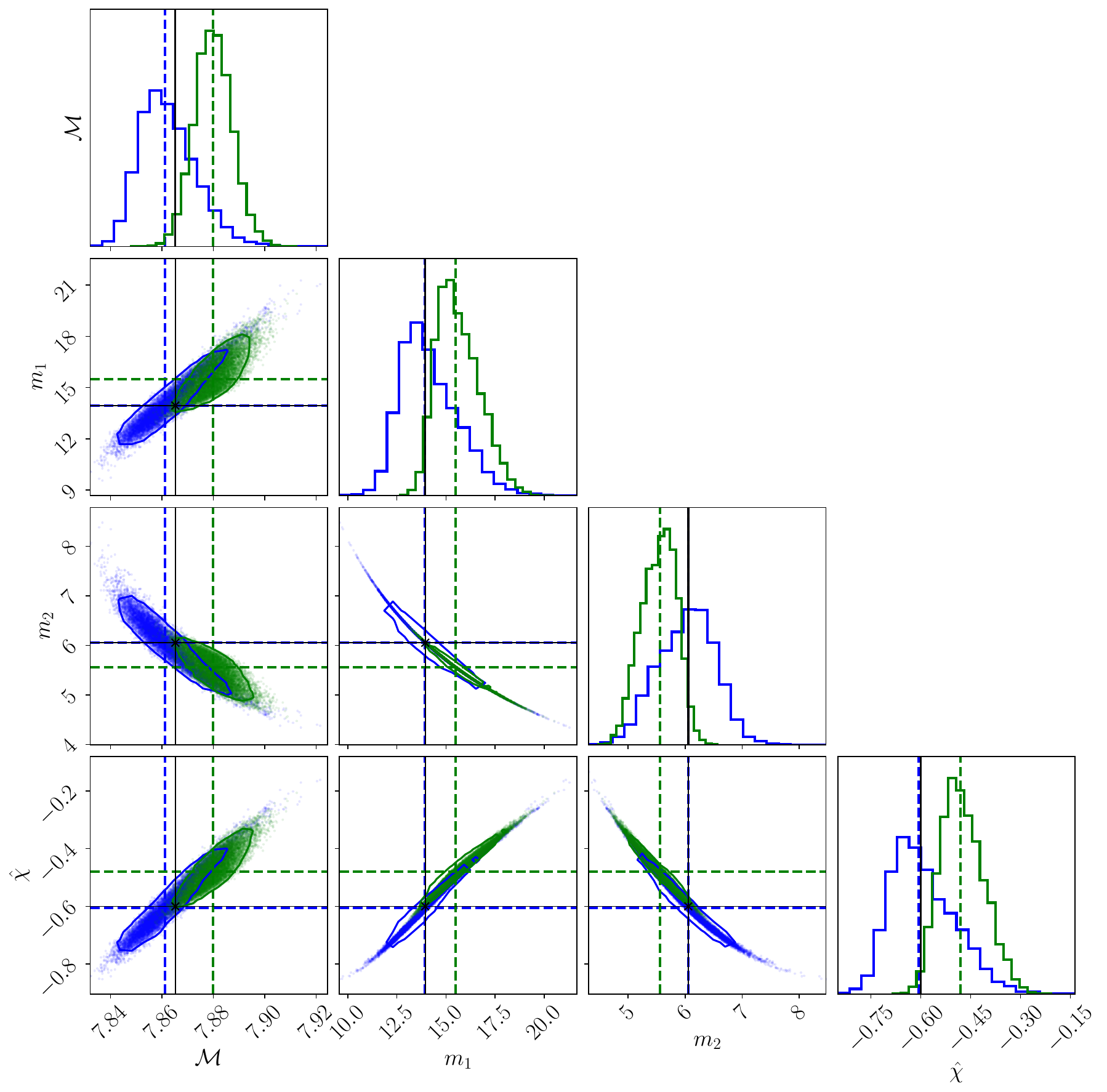}
    \caption{Corner plot showing the impact of realignment and uncertainty-marginalization on (marginalized) posterior distributions for $\mathcal{M}$, $m_1$, $m_2$, and $\hat{\chi}$. The marginalized distributions are obtained via  Eq.~\eqref{eq:marginalized_astro_distribution}. The injection (black line) is \texttt{NRHybSur3dq8}, and two recovery models are compared: the uncertainty-aware model $h(\boldsymbol{\theta}, \boldsymbol{\lambda})$  (blue histogram) and the original \texttt{IMRPhenomD} (green histogram). The dashed lines mark the respective medians.
    The relevant injection parameters are $\lbrace M, q, \chi_1,\chi_2 \rbrace= \lbrace20 M_\odot, 2.3, -0.6, -0.6\rbrace$, at an SNR of 212. The training mismatch spread is set to $\sigma_\mathrm{MM} = 10^{-4}$. Observe that, while the uncertainty-aware model shows no bias, the injection falls outside the $90\%$-credible region for the original \texttt{IMRPhenomD} model.}
    \label{fig:post_corner}
\end{figure*}

The posterior distributions obtained using the uncertainty-aware \texttt{IMRPhenomD} model, $h(\boldsymbol{\theta}, \boldsymbol{\lambda})$, with flexible fitting coefficients exhibit two key features. First, it more accurately recovers the astrophysical parameters of an injected noise-free \texttt{NRHybSur3dq8} signal than the original \texttt{IMRPhenomD} model. 
Figure~\ref{fig:post_corner} illustrates such an injection-and-recovery scenario, with injected astrophysical parameters deliberately chosen from the high mismatch (before the training) region in Fig.~\ref{fig:mismatch_validation} (specifically, at mass ratios $1 \leq q \lesssim 5$ and spin parameter $\chi_\mathrm{PN} \lesssim -0.5$) where the original \texttt{IMRPhenomD} is expected to perform poorly. We see that the uncertainty-aware model successfully recovers both masses and spin parameter, while the original \texttt{IMRPhenomD} does not.

Observe that to assess bias in the spin sector of Fig.~\ref{fig:post_corner}, we consider the normalized spin parameter\footnote{Since $\hat{\chi}$ is bounded between $-1$ and $1$, it is often preferable to plot instead of $\chi_\mathrm{PN}$.}
\begin{equation}
    \hat{\chi} = \frac{\chi_{\text{PN}}}{1 - \frac{76 \eta}{113}},
\end{equation}
rather than the individual spin components $\chi_{1,2}$. We do so because the spin parameter that enters the waveform explicitly is $\chi_\mathrm{PN}$ and extrapolating both spins from a linear combination might be beyond the capability of the aligned-spin model.

The second key feature of the posterior distributions obtained with the uncertainty-aware \texttt{IMRPhenomD} model is the following. For regions of parameter space with sufficiently high values of $\sigma_\mathrm{MM}^2$, which indicate a broader distribution $p(\boldsymbol{\lambda})$, the model produces wider marginalized posterior distributions of the astrophysical parameters, as compared to those produced by the original \texttt{IMRPhenomD} model. For the broadening to be noticeable at values of $\sigma_\mathrm{MM}$ within $10^{-3}-10^{-2}$, we may require injections with an SNR of order $10^3$, for which \texttt{bilby} is not yet fully optimized. Nonetheless, even if the impact on the parameter inference is not readily apparent at the SNRs considered in this work, we can use the mismatch distributions in Fig.~\ref{fig:mismatch_validation} to demonstrate that the prior $p(\boldsymbol{\lambda})$ is functioning as intended.

A limitation of using $p(\boldsymbol{\lambda})$ in the inference of astrophysical parameters is that it does not alleviate, but rather it intensifies the implicit dependence of the calibration procedure on the total mass $M$. Because the mismatch in Eq.~\eqref{eq:mismatch-def} carries an implicit total mass dependence (or equivalently, a dependence on the frequency range), the resulting distribution $p(\boldsymbol{\lambda})$ is effectively trained for a specific mass scale, or number of observable GW cycles. This can introduce bias when we apply $p(\boldsymbol{\lambda})$ to recover total masses for which it was not trained. For instance, the \texttt{IMRPhenomD} calibration in \texttt{ripplegw} is performed at $M = 50 M_\odot$, and when compared to a \texttt{NRHybSur3dq8} signal with $M = 10 M_\odot$, it produces mismatches of order $10^{-2}$, as seen in Fig.~\ref{fig:mismatch_validation}. Our uncertainty-aware \texttt{IMRPhenomD} model inherits this artifact. In particular, observe that it can introduce bias at SNRs $>100$ when recovering an \texttt{IMRPhenomD} injection at $(M = 50 M_\odot, q= 7.9)$ with the same \texttt{IMRPhenomD} model but equipped with a distribution $p(\boldsymbol{\lambda})$ trained on $(M = 10 M_\odot,1<q<8)$. 

Unlike the Dirac delta function prior of the original \texttt{IMRPhenomD} model, the multivariate Gaussian prior $p_\mathrm{MVG}(\boldsymbol{\lambda})$ stores a complete description of the model fit, but it can also embed unphysical correlations between astrophysical parameters. If these are activated during inference, they can lead to poor convergence of the posterior distributions, even if we inject and recover with the same model. However, we observe that in such cases the time-and-phase-marginalized -- the same settings employed in the training -- maximum likelihood point aligns well with the injection point, even though the posterior distribution does not. This leads to a possible solution for the mismatch-degrading unphysical correlations, at any mass scale. That is, one could implement a two-step inference: (i) an initial broader parameter estimation aimed at constraining the total mass, using the maximum-likelihood estimate, and (ii) a refined estimation of the same signal using a distribution for fitting coefficients $p(\boldsymbol{\lambda})$ tailored to the previous total mass estimate.
Viable alternatives to this two-step inference include extending the training to incorporate total mass as an additional dimension, or defining the training mismatch as the maximum mismatch over a range of total masses, as done in EOB model calibration \cite{Bohe:2016gbl}.

\section{Conclusions}
\label{sec:conclusions}
This paper illustrates a scalable method to propagate uncertainties of the fitting coefficients within semi-analytical GW models by sampling them from a distribution $p(\boldsymbol{\lambda})$ during estimation of astrophysical parameters. This distribution is constructed to ensure consistency with a training set of NR surrogate waveforms to a predefined mismatch resolution, meant to emulate the intrinsic uncertainty in the NR waveforms and the fitting procedure itself.
Describing the fitting coefficients through a probability distribution rather than through point estimates makes the waveform model more aware of the training waveforms used and allows for a straightforward marginalization of these uncertainties.

We applied this recipe to the late-inspiral \texttt{IMRPhenomD} phase and found that a training set of 94 \texttt{NRHybSur3dq8} surrogate waveforms is sufficient for the uncertainty-aware model to interpolate, with preset accuracy, the whole parameter space (where the surrogates are valid). The resulting probability distribution $p(\boldsymbol{\lambda})$ for the late-inspiral fitting coefficients is well described by a highly correlated multivariate Gaussian. With this, we were able to more accurately infer the astrophysical parameters of an \texttt{NRHybSur3dq8} injection when the original \texttt{IMRPhenomD} model exhibited bias.
Despite the expanded parameter space introduced by sampling over $\boldsymbol{\lambda}$, the uncertainty-aware model retains computational costs comparable to the original \texttt{IMRPhenomD}. This makes it well-suited for high-SNR parameter estimation through MCMC samplers, where a narrower posterior distribution typically leads to a lower acceptance rate, and thus, to a larger number of likelihood evaluations. 

While our method improves the consistency of semi-analytical models with NR waveforms, the calibration  remains weakly dependent on the total mass used during training, due to the definition of the mismatch. This dependence can introduce biases in same-model injection and recovery, when $p(\boldsymbol{\lambda})$ is applied to total masses significantly different from those used during training. We identified a possible solution in a two-step inference procedure. First, we constrain the total mass with a broader parameter estimation analysis. Then, we refine the analysis using a prior on the fitting coefficients suited for that mass scale. Moreover, since our analysis is restricted to the inspiral phase, no information from the merger-ringdown region was used during training. Incorporating higher frequencies is expected to reduce the mismatch sensitivity to the total mass from the early inspiral cycles. In turn, this should help prevent unphysical correlations and tails from emerging in $p(\boldsymbol{\lambda})$.

Future work could extend this approach in several directions. First, the method proposed here could be readily applied to state-of-the-art, dominant-multipole, phenomenological waveforms, like \texttt{IMRPhenomXAS}. Additionally, a more comprehensive recalibration could be achieved by fully leveraging automatic differentiation, which \texttt{ripplegw} natively supports. This would enable the joint recalibration of inspiral, merger and ringdown fitting coefficients, allowing the model to incorporate higher frequencies and possibly fit to raw NR waveforms, rather than relying solely on the semi-analytic inspiral.

Furthermore, the strong correlations observed in the inferred distribution of fitting coefficients suggest dimensionality reduction (see e.g. \cite{Saleem:2021nsb}) could be applied. If the variability in the fitting coefficients can be captured by a few principal components, then the underlying family of waveforms could be effectively reconstructed using only the most relevant degrees of freedom. This would simplify gravitational-wave inference, making parameter estimation more efficient and improving the convergence of sampling algorithms. In general, the training process itself can serve as a diagnostic tool to assess the quality of the parametrization and to identify the minimal set of components required to retain, for instance, 99\% of the waveform information.

Finally, it would be valuable to carry out a survey of systematic biases in tests of GR, mapping out the SNR and the type of astrophysical system miscalibration that leads to large biases. With this information, one could study whether the uncertainty-aware method developed here could be used to ameliorate such biases. 
Such a study would be particularly relevant for parametrized tests \cite{Li:2011cg,Li:2011vx,Agathos:2013upa}, such as the parametrized post-Einsteinian framework \cite{Yunes:2009ke, Cornish:2011ys, Chatziioannou:2012rf, Sampson:2013lpa, Moore:2021eok}, that rely on deviations from \texttt{IMRPhenomD}-like waveforms. If the base waveform used for such tests already carries calibration errors, it is essential to determine whether the inferred deviations from general relativity are genuine or artifacts of the modeling assumptions.

\section*{Acknowledgments}
We thank Rohit S. Chandramouli, Abhishek V. Joshi, Kristen E. Schumacher, Yiqi Xie, and Nijaid Arredondo for valuable suggestions. We thank Michael P\"urrer for helpful comments. 
N.Y. and S.M. acknowledge support from the Simons Foundation through Award No.~896696, the NSF through Grant No.~PHY-2207650, and NASA through Grant No.~80NSSC22K0806. 
C.-J.~H. acknowledges the support from the Nevada Center for Astrophysics, from NASA Grant No. 80NSSC23M0104, and the NSF through the Award No.~PHY-2409727.
N. C. acknowledges support from the NSF through Grant No.~PHY-2207970 and NASA through Grant No.~80NSSC24K0435. 
C.O. is supported by 
ERC Starting Grant No.~945155--GWmining, 
Cariplo Foundation Grant No.~2021-0555, 
MUR PRIN Grant No.~2022-Z9X4XS, 
MUR Grant ``Progetto Dipartimenti di Eccellenza 2023-2027'' (BiCoQ),
and the ICSC National Research Centre funded by NextGenerationEU.

Several open-source software packages were used for this work, including \texttt{numpy} \cite{harris2020array}, \texttt{matplotlib} \cite{Hunter:2007}, \texttt{scipy} \cite{2020SciPy-NMeth}, \texttt{jax} \cite{jax2018github}, \texttt{flowMC} \cite{flowmc_paper}, \texttt{ripplegw} \cite{ripple_paper}, \texttt{arviz} \cite{arviz_2019}, \texttt{corner} \cite{corner}, \texttt{bilby} \cite{bilby_paper}, \texttt{parallel-bilby} \cite{pbilby_paper}, \texttt{dynesty} \cite{2020MNRAS.493.3132S,sergey_koposov_2024_12537467}, and \texttt{singularity} \cite{singularity_paper,singularity_zenodo}.

This research was supported in part by the Illinois Computes project which is supported by the University of Illinois Urbana-Champaign and the University of Illinois System.

\bibliography{paper_arxiv_updated_after_PRD_publication}

\end{document}